\documentclass[aps,prl,reprint,amsmath,amssymb,twocolumn,superscriptaddress,notitlepage,groupedaddress]{revtex4-1}
\usepackage{graphicx}
\usepackage{dcolumn}
\usepackage{stackengine}
\usepackage[dvipsnames*,svgnames]{xcolor}
\usepackage{verbatim}
\usepackage[position=bottom,caption=false,captionskip=3pt,font=normalsize,subrefformat=simple,labelformat=simple]{subfig}

\usepackage{listings}
\usepackage{mathtools}
\usepackage{amsmath}
\usepackage{chngcntr}
\usepackage{bm}
\usepackage{url}

\usepackage[breaklinks]{hyperref}
\usepackage{breakurl}
\usepackage{tabularx}
\usepackage{soul}
\hypersetup{
colorlinks,
linkcolor={blue!100!black},
citecolor={blue},
urlcolor={blue!80!black}}
\newcommand{\fakesection}[1]{\par\refstepcounter{section}\sectionmark{#1}\addcontentsline{toc}{section}{\protect\numberline{\thesection}#1}}\def\equationautorefname~#1\null{#1\null}
\usepackage[bottom]{footmisc} 
\begin{document}
\title{Diffusive Phonons in Nongray Nanostructures}
\author{Giuseppe Romano}
\email{romanog@mit.edu}
\affiliation{Department of Mechanical Engineering, Massachusetts Institute of Technology, 77 Massachusetts Avenue, Cambridge, MA 02139, USA}
\author{Alexie M. Kolpak}
\affiliation{Department of Mechanical Engineering, Massachusetts Institute of Technology, 77 Massachusetts Avenue, Cambridge, MA 02139, USA}
\begin{abstract}
Nanostructured semiconducting materials are promising candidates for thermoelectrics due to their potential to suppress phonon transport while preserving electrical properties. Modeling phonon-boundary scattering in complex geometries is crucial for predicting materials with high conversion efficiency. However, the simultaneous presence of ballistic and diffusive phonons challenges the development of models that are both accurate and computationally tractable. Using the recently developed first-principles Boltzmann transport equation (BTE) approach, we investigate diffusive phonons in nanomaterials with wide mean-free-path (MFP) distributions.  First, we derive the short MFP limit of the suppression function, showing that it does not necessarily recover the value predicted by standard diffusive transport, challenging previous assumptions. Second, we identify a Robin type boundary condition describing diffuse surfaces within Fourier's law, extending the validity of diffusive heat transport in terms of Knudsen numbers. Finally, we use this result to develop a hybrid Fourier/BTE approach to model realistic materials, obtaining excellent agreement with experiments. These results provide insight on thermal transport in materials that are within experimental reach and open opportunities for large-scale screening of nanostructured thermoelectric materials.\end{abstract}
\maketitle

\fakesection{fake}
Due to their ability to convert heat directly into electricity, thermoelectric (TE) materials have a wide range of applications, including waste heat recovery~\cite{bell2008cooling}, wearable devices~\cite{siddique2017review}, and deep-space missions~\cite{ritz2004multi}. Widespread of TE materials is limited, however, by the simultaneous requirement for low thermal conductivity and high electrical conductivity, a condition that is rarely met in natural materials ~\cite{snyder2008complex}. Nanostructured materials overcome this limitation in that heat-carrying phonons have mean free paths MFPs ($\Lambda$) larger than the limiting dimension, $L_c$, resulting in strong thermal transport suppression~\cite{chenbook}. On the other side, electrons have MFPs that are typically as small as a few nanometers thus their size effects are mostly negligible~\cite{liao2015significant}. Notable nanostructures, including thin films~\cite{Venkatasubramanian2001}, nanowires ~\cite{hochbaum2008enhanced,boukai2008silicon}, and porous materials~\cite{song2004thermal,lee2015ballistic,Tang2010,Hopkins2011,verdier2017thermal,vega2016thermal}, show a significant suppression in thermal conductivity with respect to the bulk, holding promises for high-efficiency thermal energy conversion.\par In the case of materials with wide bulk MFP distribution, $K(\Lambda)$, the effective thermal conductivity ($\kappa_{\mathrm{eff}}$) can be conveniently computed by $\kappa_{\mathrm{eff}}/\kappa_{\mathrm{bulk}}=\int B_0(\Lambda) S(Kn)  d \Lambda$, where $\kappa_{\mathrm{bulk}}$ is the bulk thermal conductivity, $Kn=\Lambda/L_c$ is the Knudsen number, $S(Kn)$ is the material-independent, suppression function~\cite{Yang2013}, and  \begin{equation}
\begin{split}\label{Eq:1}
B_n(\Lambda)=\left[\frac{K(\Lambda)}{\Lambda^n}\right]\left[\int \left(\frac{K(\Lambda')}{\Lambda'^n} \right) d\Lambda'\right]^{-1}
\end{split}
\end{equation}is a bulk material property, which can be computed from first-principles~\cite{broido2007intrinsic}. This approach, which we refer to as the ``non-interacting model,'' treats phonons with different MFPs separately, with the diffusive regime (i.e., for short $Kns$) as described by standard Fourier's law; on the other hand, the suppression function associated with ballistic phonons (i.e., with large $Kns$) is $S(Kn)\approx Kn^{-1}$, arising from the MFPs in the nanostructure being limited b $L_c$. The non-interacting model is exact only for simple geometries, such as thin films, where analytically expression are available~\cite{fuchs1938conductivity,film2}. However, for more complex geometries, one has to solve the space-dependent BTE, which gives $\kappa_{\mathrm{eff}}/\kappa_{\mathrm{bulk}}=\int B_0(\Lambda) \overline{S}(\Lambda,\Omega) d\Lambda $, where $\overline{f}=\left(\int_{4\pi} f(\Omega) d\Omega \right)^{-1}$ is an angular average along the solid angle $\Omega$, representing phonon direction. Generally, $S(\Lambda,\Omega)$, the ``directional'' suppression function~\cite{romano2016directional}, depends on $K(\Lambda)$ itself~\cite{romano2015};  thus the phonon suppression at a given $\Lambda$ depends on the whole bulk spectrum. Consequently, the notion of diffusive and ballistic regimes has to be revisited in order to incorporate the coupling between phonons at different MFPs.\par  
\begin{figure*}

\subfloat[\label{Fig:10a}]
{\includegraphics[width=0.48\textwidth]{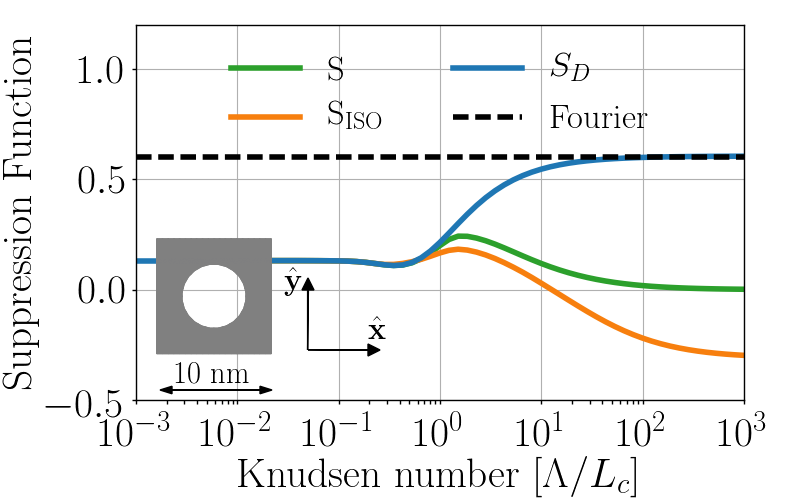}}
\hfill
\subfloat[\label{Fig:10b}]
{\includegraphics[width=0.48\textwidth]{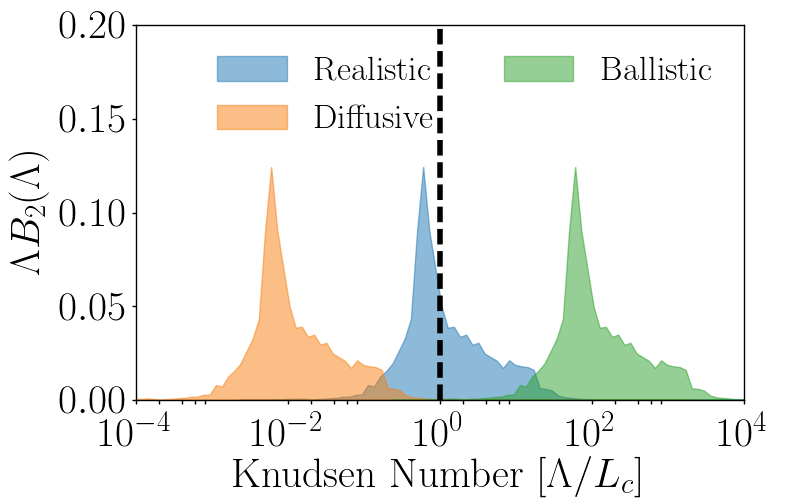}}
\quad
\subfloat[\label{Fig:10c}]
{\includegraphics[width=0.48\textwidth]{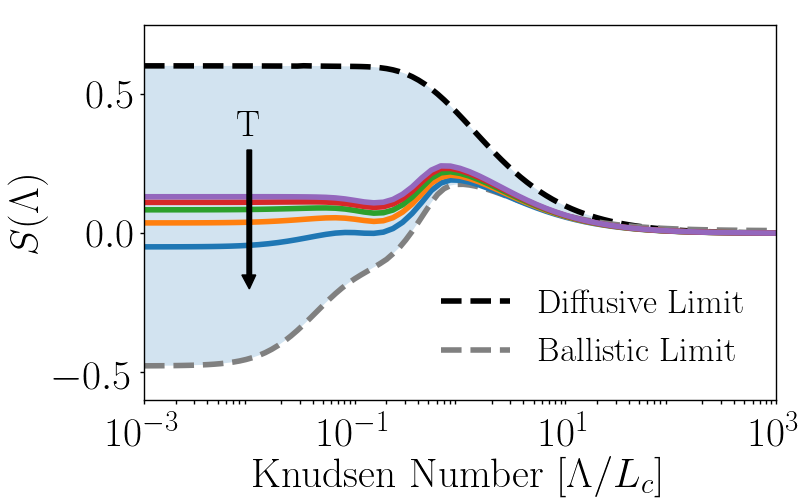}}
\hfill
\subfloat[\label{Fig:10d}]
{\includegraphics[width=0.48\textwidth]{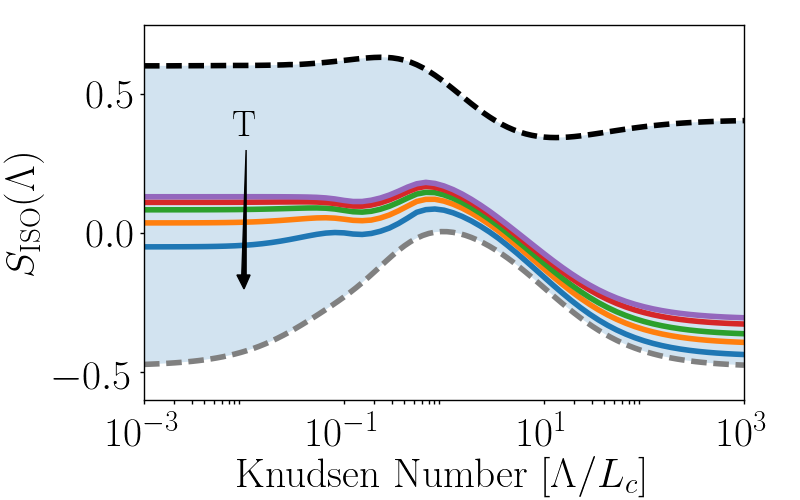}}
\caption{(a) For short $Kns$, the suppression function, $\overline{S}(\Lambda)$, reaches a plateau that is significant lower than that calculated by the standard Fourier's law. Up to $Kn \approx 1$, the isotropic suppression function, $S_{\mathrm{ISO}}(\Lambda)\approx \overline{S}(\Lambda)$, because the phonon distributions are isotropic. The diffusive suppression function, $S_D(\Lambda)$, reveals the breakdown of Fourier's law for $Kn > 1$. In the inset, the unit cell including a single circular pore and with periodicity L = 10 nm. (b) The coefficients $B_2(\Lambda)$ for a realistic, diffusive and ballistic materials. The dotted line represents the characteristic length, $L_c$. (c) $S(\Lambda)$ and (d) $S_{\mathrm{ISO}}(\Lambda)$ for the case of ballistic and diffusive materials. Realistic materials lie in the shaded regions. The curves for Si at T = 150, 200, 250, and 300 K are shown for comparison.}
\end{figure*}
Using our recently developed solver for the space-dependent BTE, we provide a framework to understand the effect of ballistic phonons on heat diffusion in nanostructures. First, we provide an analytical expression for the small-MFP limit of the suppression function, demonstrating a significant departure from the non-interacting model. We also show that the non-interacting model is the special case when \textit{all} the MFPs of heat-carrying phonons are smaller than $L_c$. Second, we investigate the effect of large-$Kn$ phonons on the diffusive thermal flux along the wall of the pores, identifying a Robin type boundary condition that, essentially, extends the range of validity of Fourier's law.  Finally, using these two findings we implement a hybrid Fourier/BTE model to calculate $\kappa_{\mathrm{eff}}$ in realistic porous samples, obtaining excellent agreement with experiments. Our work extends our knowledge of heat transport in nanostructured materials and provides insights for \textit{ab initio}, multiscale thermal conductivity calculations.\par We model phonon transport via the MFP-dependent BTE~\cite{romano2015}  \begin{equation}
\begin{split}\label{Eq:2}
\Lambda \mathbf{\hat{s}}(\Omega) \cdot \nabla T(\mathbf{r},\Lambda,\Omega) = T_L(\mathbf{r})-T(\mathbf{r},\Lambda,\Omega),
\end{split}
\end{equation}where $T(\mathbf{r},\Lambda,\Omega)$ is an effective, space-dependent temperature associated with phonons with MFP $\Lambda$ and direction $\mathbf{\hat{s}}(\Omega)$ denoted by $\Omega$; the term $T_L(\mathbf{r})$ is an effective lattice temperature, obtained by  \begin{equation}
\begin{split}\label{Eq:3}
T_L(\mathbf{r})=\int B_2(\Lambda) \overline{T}(\mathbf{r},\Lambda,\Omega) d\Lambda.
\end{split}
\end{equation}Within this formalism, the normalized thermal flux is $\mathbf{J}(\mathbf{r},\Lambda,\Omega) = B_1(\Lambda)T(\mathbf{r},\Lambda,\Omega)\mathbf{\hat{s}}(\Omega)$~\cite{romano2015}, where we used the scaling factor $\left[\int K(\Lambda)/\Lambda d\Lambda \right]^{-1}$. For simplicity, when unambiguous, we will drop the space and angular dependencies from the notation.\par  
\begin{figure*}[!ht]

\subfloat[\label{Fig:20a}]
{\includegraphics[width=0.48\textwidth]{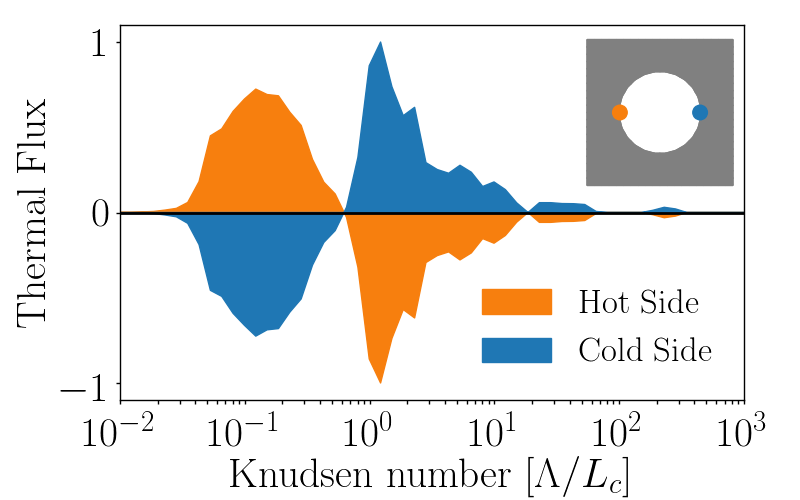}}
\hfill
\subfloat[\label{Fig:20b}]
{\includegraphics[width=0.48\textwidth]{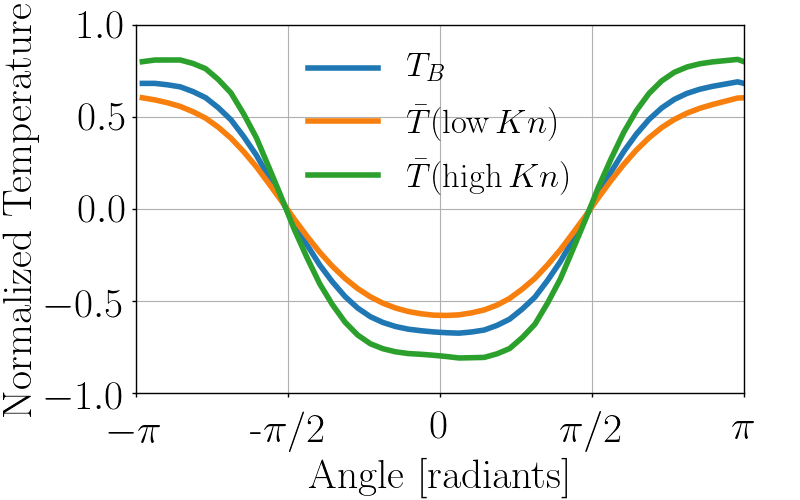}}
\caption{(a) Normal thermal flux for different $Kns$ at the hot and cold sides of the pore. (b) Temperature profile around the boundary of the pore for $T_B$, as calculated by Eq.~(\autoref{Eq:1003}), as well as for high and low $Kn$ phonons. The angles $\phi = -\pi$ and $\phi=0$ coincide with the directions $\mathbf{\hat{x}}$ and $-\mathbf{\hat{x}}$.}
\end{figure*}
We first solve Eqs.~(\autoref{Eq:2}-\autoref{Eq:3}) on ordered nanoporous Si with infinite thickness. As shown in Fig.~\subref*{Fig:10a}, we consider a two-dimensional unit-cell containing one circular pore and apply periodic boundary conditions both along $\mathbf{\hat{x}}$ and $\mathbf{\hat{y}}$. We consider the case with porosity is $\varphi=0.25$ and periodicity L = 10 nm. The walls of the pores are assumed to scatter phonons diffusively, a condition that translates into the following temperature imposed to outgoing phonons~\cite{romano2016temperature}  \begin{equation}
\begin{split}\label{Eq:1003}
T_B = \int B_1(\Lambda) g(\Lambda) d\Lambda  
\end{split}
\end{equation}where $g(\Lambda)$ is the average flux of incoming phonons, given by $g(\Lambda)=\left[<\mathbf{\hat{s}}\cdot\mathbf{\hat{n}}>_+ \right]^{-1}<T(\Lambda)\mathbf{\hat{s}}\cdot\mathbf{\hat{n}}>_+$. The notation $<f>_+$ stands for an angular average for the emisphere where $\mathbf{\hat{s}}\cdot\mathbf{\hat{n}}>0$. Heat flux is enforced by applying a difference of temperature $\Delta T =$ 1K between the hot and cold contacts, as illustrated in Fig.~\subref*{Fig:10a}. Once Eqs.~(\autoref{Eq:2}-\autoref{Eq:3}) are solved iteratively, we compute the directional suppression function~\cite{romano2016directional}  \begin{equation}
\begin{split}\label{Eq:5}
S(\Lambda,\Omega) = -3\frac{L}{\Delta T}\mathbf{\hat{s}} \otimes \mathbf{\hat{s}} \nabla <T(\Lambda)>_{A_{\mathrm{hot}}}\cdot \mathbf{\hat{n}},
\end{split}
\end{equation}where $<f>_{A_{\mathrm{hot}}}=\left(A_{\mathrm{hot}} \right)^{-1}\int_{A_{\mathrm{hot}}} f dS$ is a spatial average along the hot contact, denoted by $A_{\mathrm{hot}}$. In agreement with previous results~\cite{romano2016temperature}, $\kappa_{\mathrm{eff}} \approx $  6 Wm$^{-1}$ K$^{-1}$, significantly lower than the bulk value $\kappa_{\mathrm{bulk}} = \int K(\Lambda)d\Lambda \approx $ 153 Wm$^{-1}$ K$^{-1}$~\cite{li2014shengbte}. The angularly averaged suppression function $\overline{S}(\Lambda)$, simply referred to as the suppression function, is shown in Fig.~\subref*{Fig:10a}. We note that for large $Kns$, $\overline{S}(\Lambda) \propto  L_c/\Lambda $ in accordance with the ballistic regime, whereas suppression of phonons with short $Kns$ is constant with MFP until approaching the quasi-ballistic regime, i.e. for $Kn \approx $ 1. The latter region is the focus of our study.\par For small $Kns$, phonon distributions are isotropic and can be expanded to first-order spherical harmonics $T(\Lambda)\approx \overline{T}(\Lambda) - \Lambda \mathbf{\hat{s}}\cdot\nabla\overline{T}(\Lambda)$, which, when combined with Eq.~(\autoref{Eq:5}) and after an angular average, leads to  \begin{equation}
\begin{split}\label{Eq:1005}
\overline{S}_{\mathrm{ISO}}(\Lambda) = -\frac{L}{\Delta T}<\nabla \overline{T}(\Lambda)\cdot\mathbf{\hat{n}}>_{A_{\mathrm{hot}}},
\end{split}
\end{equation}where ISO stands for ``isotropic,'' and we used $\overline{\mathbf{\hat{s}}\otimes \mathbf{\hat{s}} } = (1/3) \delta_{ij}$. Heat transport in the short MFP region is calculated by including this expansion in Eq.~(\autoref{Eq:2}), obtaining~\cite{loy2013fast}  \begin{equation}
\begin{split}\label{Eq:1002}
\frac{\Lambda^2}{3}\nabla^2 \overline{T}(\Lambda) = \overline{T}(\Lambda)-T_L.
\end{split}
\end{equation}Equation~(\autoref{Eq:1002}) is the diffusive heat conduction equation with effective heat sources arising from the coupling between phonons with different MFPs.  
\begin{figure*}

\subfloat[\label{Fig:30a}]
{\includegraphics[width=0.48\textwidth]{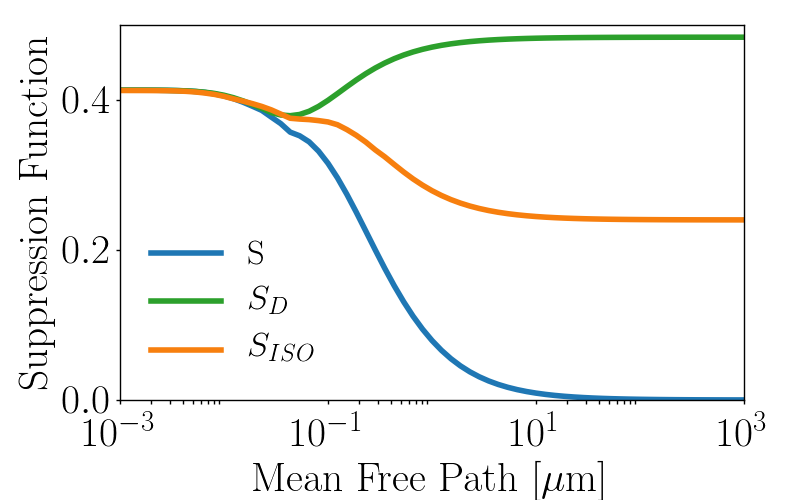}}
\hfill
\subfloat[\label{Fig:30b}]
{\includegraphics[width=0.48\textwidth]{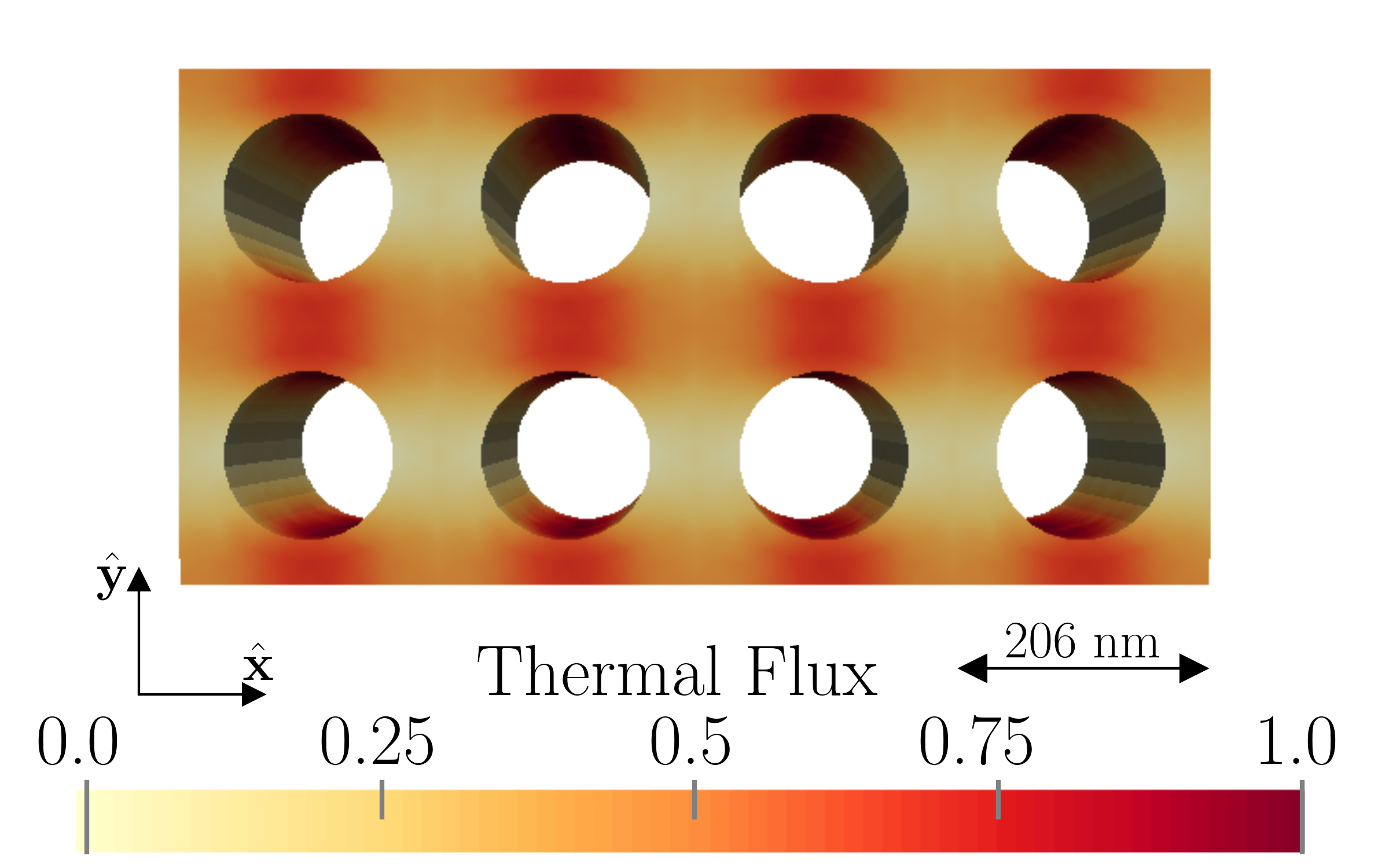}}
\quad
\subfloat[\label{Fig:30c}]
{\includegraphics[width=0.48\textwidth]{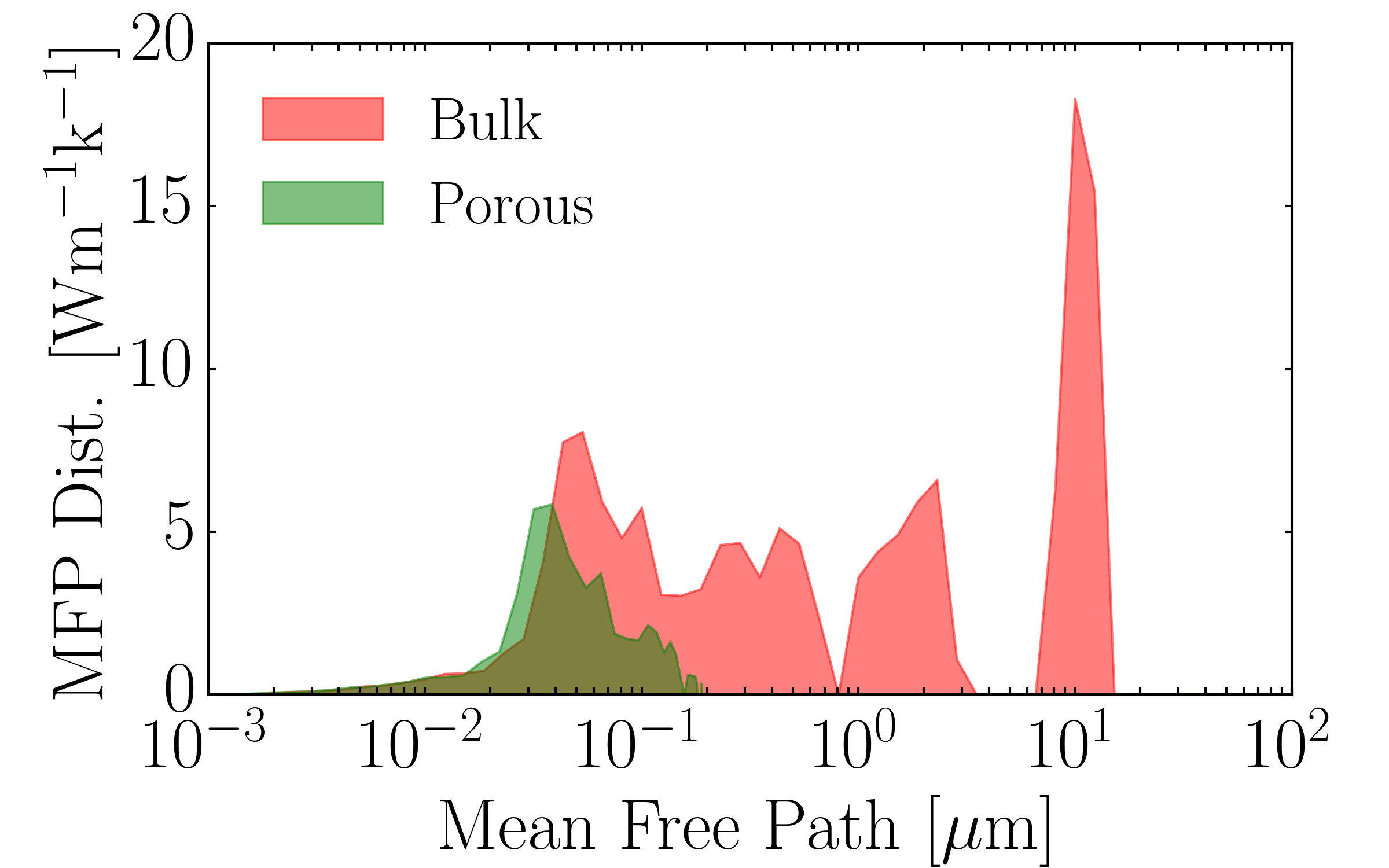}}
\hfill
\subfloat[\label{Fig:30d}]
{\includegraphics[width=0.48\textwidth]{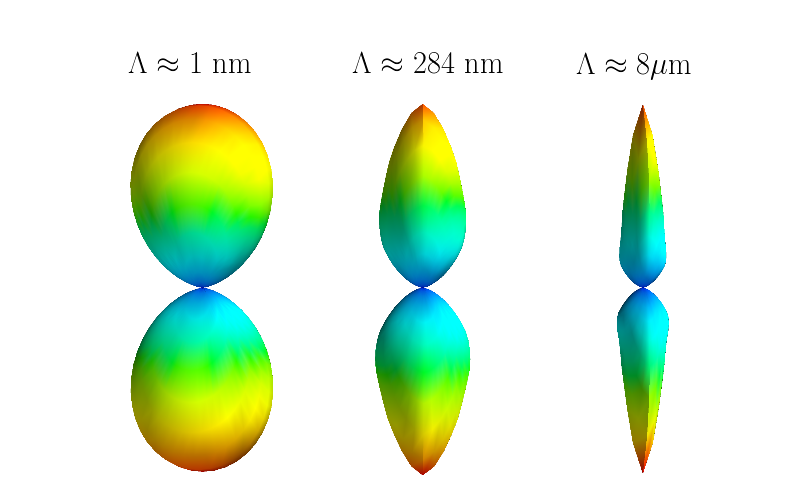}}
\caption{(a) The suppression function, $S(\Lambda)$, the isotropic suppression function, $S_{ISO}(\Lambda)$, and the diffusive suppression function, $S_D(\Lambda)$, for the structure presented in~\cite{vega2016thermal}. (b) The thermal flux profile. As expected, most of the heat travel along the space between pores along the temperature gradient. (b) The suppression function, $S(\Lambda)$, the isotropic suppression function, $S_{ISO}(\Lambda)$, and the diffusive suppression function for the structure presented in~\cite{vega2016thermal}. (c) The MFP distribution in the porous material. The maximum allowed MFP is around 200 nm, while in bulk phonons have MFPs up to tens of microns. (d) The directional phonon suppression function for different $Kns$. As heat becomes ballistic (i.e., for high $Kns$), phonons acquire peaked directionality.}
\end{figure*}
For $\Lambda \rightarrow 0$, Eq.~(\autoref{Eq:1002}) simplifies to $\overline{T}(0)=T_L$, which, after using Eqs.~(\autoref{Eq:3}-\autoref{Eq:1005}), gives  \begin{equation}
\begin{split}\label{Eq:8}
\overline{S}(0) = \int_{0}^{\infty} B_2(\Lambda) S_{\mathrm{ISO}}(\Lambda)d\Lambda,
\end{split}
\end{equation}From Eq.~(\autoref{Eq:8}), the first key result of this paper, we note that $\overline{S}(0)$ depends on the entire bulk MFP distribution, embodying the effect of ballistic phonons on diffusive heat.\par The upper bound of $\overline{S}(0)$ is evaluated by introducing the concept of ``diffusive materials,'' i.e. a material where all the MFPs for which $B_2(\Lambda)$ is significant are much smaller than $L_c$, as depicted in Fig.~\subref*{Fig:10b}. Under this condition, $T_L = \overline{T}(0)$ and Eq.~(\autoref{Eq:1002}) becomes the Laplacian  $\nabla^2 \overline{T}(0) = 0$. Moreover, the boundary temperature becomes $T_B=\overline{T}(0)$. Then, $\overline{S}(0)$ is given by Eq.~(\autoref{Eq:1005}). For the case of ordered circular pores, this model gives $\overline{S}(0)\approx \left(1-\varphi\right)/\left(1 + \varphi\right) = 0.6 $, as illustrated in Fig.~\subref*{Fig:10a}. This value coincides with that of the non-interacting model and can be seen as the diffusive limit of a diffusive material.\par A lower limit to $\overline{S}(0)$ can be achieved in the case of a ``ballistic material,'' namely when the MFPs contributing to $B_2(\Lambda)$ are much larger than $L_c$, as shown in Fig.~\subref*{Fig:10b}. Within this regime, $T_B = g(\infty)$ and $\overline{S}(0)=-\left(L/\Delta T \right)<\nabla \overline{T}(\infty)\cdot\mathbf{\hat{n}}>_{A_{\mathrm{hot}}}\approx -0.5$, with $\overline{T}(\infty)$ computed by Eq.~(\autoref{Eq:2}). Although a negative suppression function is counterintuitive, note that the actual MFPs in the nanostructure $\Lambda_{\mathrm{nano}} = \overline{S}(\Lambda)\Lambda$ are still positive. Again, $\overline{S}_{\mathrm{ISO}}(0)=\overline{S}(0)$ because for short MFPs the phonon distributions are isotropic. Furthermore, $S_{\mathrm{ISO}}(0)= S_{\mathrm{ISO}}(\infty)$, as demonstrated by simply including $\overline{T}(\infty)$ in Eq.~(\autoref{Eq:5}), and shown in Fig.~\subref*{Fig:10d}. In the case of realistic materials $\overline{S}(\Lambda)$ and $S_{\mathrm{ISO}}(\Lambda)$ fall in between the diffusive and ballistic material limits, depending on the MFPs contributing to $B_2(\Lambda)$ with respect to $L_c$. In Figs.~\subref*{Fig:10c} and~\subref*{Fig:10d}, we report $\overline{S}(\Lambda)$ and $S_{\mathrm{ISO}}(\Lambda)$, respectively, for different temperatures. We note that both functions decrease with temperature, as the bulk MFPs become larger ~\cite{romano2016temperature}, resulting in a shift of $B_2(\Lambda)$ toward higher MFPs.\par We now analyze the effect of ballistic phonons on the boundary conditions along the boundary of the pore, within the diffusive regime. The condition imposed on phonons leaving the boundary, exemplified by Eq.~(\autoref{Eq:1003}), translates into the following expression for the angularly averaged, normal thermal flux  \begin{equation}
\begin{split}\label{Eq:9}
\overline{\mathbf{J}}(\Lambda)\cdot\mathbf{\hat{n}}=\frac{1}{4}B_1(\Lambda)\left[g(\Lambda) - \int B_1(\Lambda') g(\Lambda') d\Lambda'\right],
\end{split}
\end{equation}where we used $<\mathbf{\hat{n}}\cdot\mathbf{\hat{s}}>_+=(1/4)$. The first and second terms in the parenthesis of Eq.~(\autoref{Eq:9}) are related to the incoming and outgoing phonons, respectively, with respect to the boundary of the pore. To understand the power balance along the diffuse surface of the pore, we note that large $Kn$ phonons tend to accumulate at the hot side of the pore wall~\cite{jeng2008modeling}, resulting in higher value of $g(\Lambda)$ with respect to diffusive phonons. Furthermore, as $T_B$ is a weighted average of $g(\Lambda)$, we have $g(\Lambda>L_c)> T_b >g(\Lambda<L_c)$, as illustrated in Fig.~\subref*{Fig:20b}. Consequently, according to Eq.~(\autoref{Eq:9}), the normal flux is positive for small $Kns$ and negative for ballistic phonons [see Fig.~\subref*{Fig:20a}]. The transition value is close to $1$ and depends on $B_1(\Lambda)$. The normal flux at the cold side of the pore has the opposite trend.\par To derive an approximation to Eq.~(\autoref{Eq:9}) for short $Kns$, we first note that, within the diffusive regime, at a point right before the wall of the pore, heat flux is $\overline{\mathbf{J}}(\Lambda)= -B_1(\Lambda)\Lambda /3 \nabla \overline{T}(\Lambda)$. Then, we expand $g(\Lambda)$ up to its first harmonics, i.e., $g(\Lambda)=\overline{T}(\Lambda)-(2/3)\Lambda \nabla \overline {T}(\Lambda)\cdot\mathbf{\hat{n}}$, where we used $<\mathbf{\hat{s}}\otimes\mathbf{\hat{s}}>_+=(1/6)\delta_{ij}$. After combining these results, Eq.~(\autoref{Eq:9}) becomes  \begin{equation}
\begin{split}\label{Eq:1001}
\overline{\mathbf{J}}(\Lambda)\cdot \mathbf{\hat{n}}=\frac{1}{2} B_1(\Lambda)\left[\overline{T}(\Lambda)-T_B \right],
\end{split}
\end{equation}a typical Robin boundary condition for heat flux, with boundary conductance $(1/2)B_1(\Lambda)$. Equation~(\autoref{Eq:1001}) constitutes the second key result of this paper. In Fig.~\subref*{Fig:10a} we plot $S_D(\Lambda)$, the diffusive suppression function, computed by Eq.~(\autoref{Eq:1005}) but with $\overline{T}(\Lambda)$ calculated with Eqs.~(\autoref{Eq:1002}-\autoref{Eq:1001}) for the whole range of MFPs. We note a deviation from $\overline{S}(\Lambda)$ around $Kn \approx 1$.\par In this last part, we calculate the thermal conductivity of a recently fabricated porous Si membrane in which strong phonon size effects were observed ~\cite{vega2016thermal}. The sample is 340 nm thick and has circular pores with diameter of 135 nm and periodicity of L = 206 nm. The nominal porosity is $\phi \approx $ 0.337. Although the solver for Eq.~(\autoref{Eq:2}) has been conveniently parallelized, computing phonon transport in such a large simulation domain can become cumbersome. In particular, the computational bottleneck arises from the need to solve the BTE for a wide spectrum using the same space discretization. To this end, we devise a hybrid Fourier/BTE computational model that solves the BTE for long $Kns$ phonons and Fourier's law [by means of Eqs.~(\autoref{Eq:1002}-\autoref{Eq:1001})] for phonons with small $Kn$. To uniquely define the MFP delimiting the two regions, we first solve both the BTE and Fourier model for decreasing MFP starting from the highest $\Lambda$ in the bulk MFP distribution. Then, when the resulting suppression functions [shown in Fig.~\subref*{Fig:30a}] converge within 1$\%$, only the latter is used until a plateau for small $Kns$ is obtained. In this case, the BTE is solved only for about 50 $\%$ of the spectrum, allowing accurate phonon transport simulations within reasonable computational time.\par  The computed thermal conductivity is 16.8 Wm$^{-1}$K$^{-1}$, within 3 $\%$ the experimental value~\cite{vega2016thermal}. Figure~\subref*{Fig:30b}, we plot the normalized thermal flux. As the figure shows, similarly to~\cite{romano2015}, most of the flux is concentrated along the space between pores, a typical signature of phonon size effects. Figure~\subref*{Fig:30c}shows the MFP distribution $K(\Lambda S(\Lambda))S(\Lambda)$ in the membrane. We note that the maximum allowed MFP is around 200 nm, a value between the sample thickness and the pore-pore distance. This combination of characteristic lengths results in three-dimensional phonon size effects. This finding is better explained by the directional suppression functions shown in Fig.~\subref*{Fig:30d}. We implicitly describe $S(\Lambda,\Omega)$ by the equation $\mathbf{r}-S(\Lambda,\Omega)\mathbf{\hat{s}}(\Omega)=0$~\cite{romano2016directional} and we chose $\Lambda \approx$ = 1 nm, 200 nm and 8 $\mu$m. As the MFP increases, $S(\Lambda,\Omega)$ becomes peaked and shrinks in both $\mathbf{\hat{x}}$ and $\mathbf{\hat{z}}$ directions. \par In summary, using first-principles calculations and the BTE, we have rivisited the diffusive regime in nongray nanostructured materials. In particular, we investigated the effect of long-$Kn$ phonons on heat diffusion, deriving an analytical expression for the short-Kn limit of the suppression function and a Robin type boundary condition for thermal flux normal to the pore boundaries. Finally, we developed a hybrid Fourier/BTE model to calculate the thermal conductivity in realistic materials, finding excellent agreement with experiments. These findings refine the concept of diffusive transport in nongray materials and pave the way for accurate yet inexpensive modeling of nanostructured materials for thermoelectric applications. \section*{Acknowledgements}
Research supported as part of the Solid-State Solar-Thermal Energy Conversion Center (S3TEC), an Energy Frontier Research Center funded by the US Department of Energy (DOE), Office of Science, Basic Energy Sciences (BES), under Award DESC0001.
\end{document}